\pdfoutput=1
\RequirePackage{amsmath}
\documentclass[aps,pra,twocolumn,amsmath,amssymb,superscriptaddress,longbibliography]{revtex4-1}

\usepackage{amssymb}
\usepackage{bm}
\usepackage{amsmath}
\usepackage{mathtools}
\usepackage{siunitx}
\usepackage{graphicx}
\usepackage{balance}
\usepackage{mathptmx}

\usepackage{booktabs}
\usepackage{bbold}
\usepackage{xcolor}

\renewcommand{\vec}[1] {{\bm{#1}}}
\newcommand{\la}{\lambda}
\newcommand{\ham}{\mathcal{H}}
\newcommand{\ce}[1] {$\mathrm{#1}$}
\newcommand{\abs}[1] {{\lvert}#1{\rvert}}
\newcommand{\bra}[1] {\langle #1{\rvert}}
\newcommand{\ket}[1] {\lvert #1{\rangle}}
\DeclareMathOperator{\sign}{sign}

\DeclareSIUnit\rydberg{Ry}

\begin{document}

\title[Persistence of symmetry-protected Dirac points at the surface of the topological crystalline insulator SnTe upon impurity doping]
{Persistence of symmetry-protected Dirac points at the surface of the topological crystalline insulator SnTe upon impurity doping}

\author{Olga Arroyo-Gasc\'{o}n}
\email{o.arroyo.gascon@csic.es}
\affiliation{Instituto de Ciencia de Materiales de Madrid, Consejo Superior de Investigaciones Cient\'{\i}ficas, C/ Sor Juana In\'es de la Cruz 3, 28049 Madrid, Spain}
\affiliation{GISC, Departamento de F\'{\i}sica de Materiales, Universidad Complutense, E-28040 Madrid, Spain}
\author{Yuriko Baba}
\affiliation{GISC, Departamento de F\'{\i}sica de Materiales, Universidad Complutense, E-28040 Madrid, Spain}
\author{Jorge I. Cerd\'{a}}
\email{Deceased.}
\affiliation{Instituto de Ciencia de Materiales de Madrid, Consejo Superior de Investigaciones Cient\'{\i}ficas, C/ Sor Juana In\'es de la Cruz 3, 28049 Madrid, Spain}
\author{Oscar de Abril}
\affiliation{Departamento de Estructuras y F\'{\i}sica de Edificaci\'{o}n, Universidad Polit\'{e}cnica de Madrid, E--28031 Madrid, Spain}
\author{Ruth Mart\'{i}nez}
\affiliation{GISC, Departamento de F\'{\i}sica de Materiales, Universidad Complutense, E-28040 Madrid, Spain}
\author{Francisco Dom\'{i}nguez-Adame}
\affiliation{GISC, Departamento de F\'{\i}sica de Materiales, Universidad Complutense, E-28040 Madrid, Spain}
\author{Leonor Chico}
\affiliation{GISC, Departamento de F\'{\i}sica de Materiales, Universidad Complutense, E-28040 Madrid, Spain}
\affiliation{Instituto de Ciencia de Materiales de Madrid, Consejo Superior de Investigaciones Cient\'{\i}ficas, C/ Sor Juana In\'es de la Cruz 3, 28049 Madrid, Spain}

\date{\today}

\begin{abstract}
We investigate the effect of a non-magnetic donor impurity located at the surface of the SnTe topological crystalline insulator. In particular, the changes on the surface states due to a Sb impurity atom are analyzed by means of \textit{ab-initio} simulations 
of pristine and impurity-doped SnTe. Both semi-infinite and slab geometries are considered within the first-principles approach. Furthermore, minimal and Green's function continuum models are proposed with the same goal. 
We find that the Dirac cones are shifted down in energy upon doping; this shift strongly depends on the position of the impurity with respect to the surface. In addition, we observe that the width of the impurity band presents an even-odd behavior by varying the position of the impurity.  
This behavior is related to the position of the nodes of the wave function with respect to the surface, and 
hence it is a manifestation of confinement effects.
We compare slab and semi-infinite geometries within the \textit{ab-initio} approach, demonstrating that the surface states remain gapless and their spin textures are unaltered in the doped semi-infinite system. 
In the slab geometry, a gap opens due to hybridization of the states localized at opposite surfaces. Finally, by means of a continuum model, we extrapolate our results to arbitrary positions of the impurity, clearly showing a non-monotonic behavior of the Dirac cone.
\end{abstract}

\maketitle


\section{\label{sec:level1} Introduction \protect\\} 

The discovery of topological insulators (TIs) in the past decade has fueled great interest in Dirac matter. These materials are insulating in the bulk, but host protected electronic states in their boundaries; namely, surface or edge modes in the three-dimensional or two-dimensional case, respectively~\cite{Hasan2010,Qi2011}. Such gapless boundary states are protected by time-reversal-symmetry; they have a Dirac spectrum and show spin-momentum locking, a promising property for their application in spintronic devices and quantum information processing~\cite{Manchon2015}. TIs are successfully characterized by the $\mathbb{Z}_2$ topological invariant, which is related to the number of gapless Kramers pairs in their boundaries. In fact, they show an odd number of Dirac cones at their surfaces. The discovery of topological insulators has led to the recognition that symmetry-protected topological phases of matter are more abundant and ubiquitous than expected~\cite{Ando2015, Wieder2021}. 

One of the families within the broad class of symmetry-protected matter are topological crystalline insulators (TCIs). Unlike $\mathbb{Z}_2$ insulators, TCIs are protected by crystal symmetries and present an even number of Dirac cones in their surfaces, so that their $\mathbb{Z}_2$ index is zero. These symmetries include rotations, reflections, or even glide planes. The first TCIs discovered were IV-VI semiconductors, such as SnTe and Pb$_{1-x}$Sn$_x$Te ternary alloys, which are protected by mirror symmetries~\cite{Hsieh2012SnTe,Dziawa2012,Tanaka2012,xu2012observation}. When a mirror symmetry is responsible for the topological protection, the relevant topological invariant is the mirror Chern number $n_M$. There are two types of surface states in TCIs, depending on the location of the Dirac points, with different properties. Type-I surface states have their Dirac points at time-reversal invariant momenta, whereas in type-II TCIs the Dirac cones are displaced from such time-reversal invariant k-points \cite{Liu2013SnTe}.

\begin{figure}[h!]
    \centering
    \includegraphics[width=0.48\textwidth]{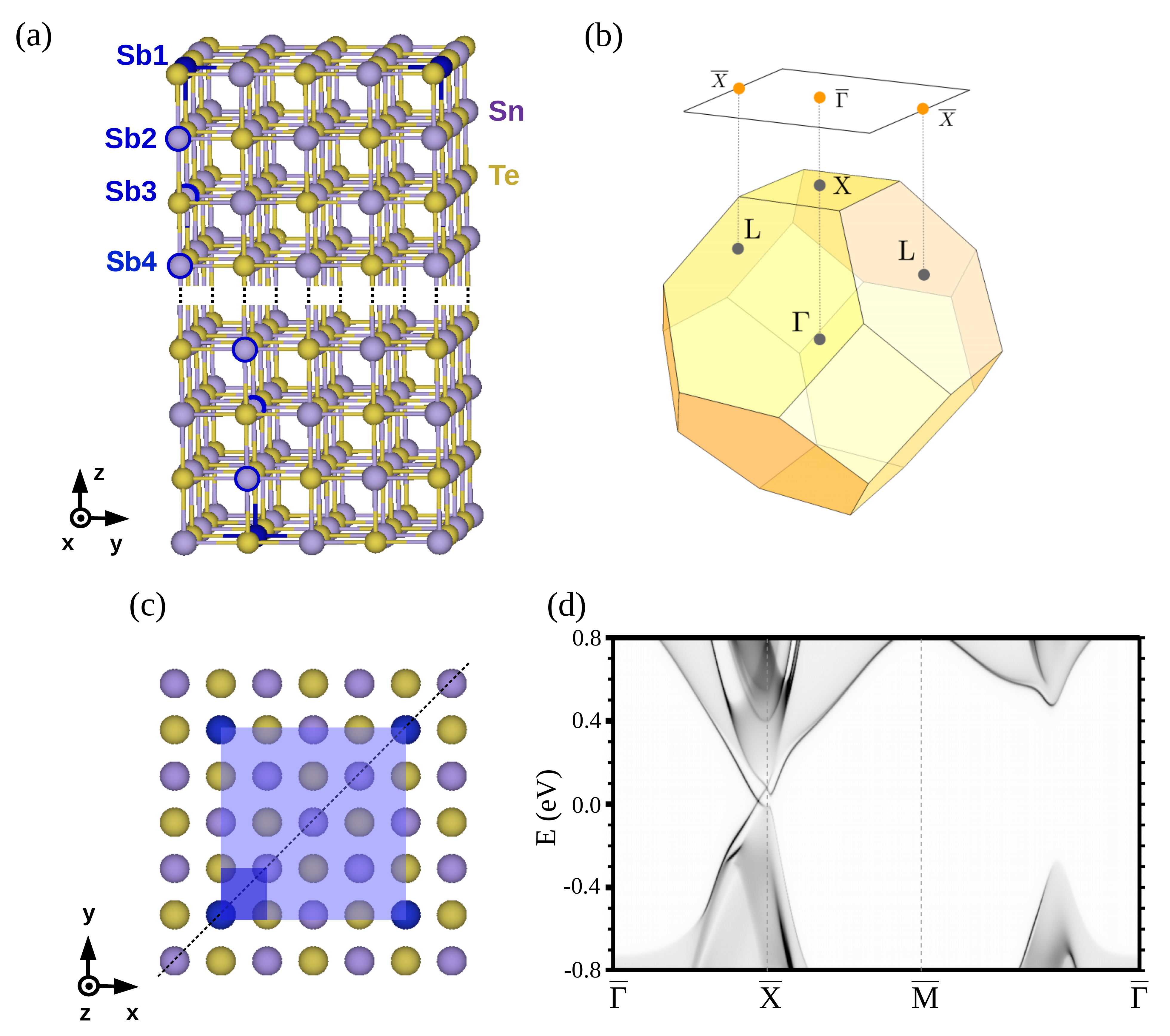}
    \caption{(a) SnTe 8-atomic-layer slab unit cell. Sb impurities are highlighted in blue: in the Sb1 case, the whole atoms are colored; in the Sb2, Sb3 and Sb4 case, the impurity atoms are circled. Higher slabs can be constructed adding new atomic layers in the dotted region. A $\SI{16}{\angstrom}$-wide vacuum space was considered in the $z$-direction in order to avoid interaction among the slabs during the first-principles calculations. (b) Face-centered cubic Brillouin zone and projected (001) plane. (c)~Top view of the c$(4\times4)$ (light blue) and $1\times1$ (dark blue) unit cells. The mirror symmetry plane is represented by a dotted line. (d)~SnTe $(001)$ surface band structure of the semi-infinite system.}
    \label{fig1}
\end{figure}

SnTe has a rocksalt crystalline structure with small bandgaps at the $L$ points of the Brillouin zone (BZ), where the conduction and valence bands are inverted \cite{Hsieh2012SnTe,Ye2015}. The mirror plane that protects these topological states is $(110)$ [see Figure \ref{fig1}(c)], so that  an even number of robust Dirac cones appear at the $( 001 )$, $( 110 )$ and $( 111 )$ crystal surfaces, which are symmetric with respect to the $( 110 )$ mirror reflection. For the (111) termination, surface states are of type-I, whereas (001) and (110) surface states are identified as type-II. For the $[001]$ surface termination, which is of type II, two equivalent $L$ points of the BZ are projected in the $\Gamma L_1L_2$ plane to the same $\bar{X}$ point, as depicted in Figure~\ref{fig1}(b). Hence, two parent Dirac cones coexist at $\bar{X}$, and their  interaction gives rise to two pairs of massless Dirac fermions deviated from $\bar{X}$, while the high-symmetry point is found to be gapped \cite{Liu2013SnTe} [see Figure~\ref{fig1}(d)]. SnTe undergoes a crystalline phase transition at low temperatures (below 100 K) to a rhombohedral lattice. Such distortion has a small effect in the bulk bands, but destroys the surface Dirac cones, as any other perturbation that breaks the mirror  $(110)$ symmetry would do~\cite{Okada2013}. Interestingly, within this rhombohedral phase, SnTe has been recently shown to be a higher-order TI, showing hinge states along specific directions~\cite{Schindler2018}. 

On the other hand, PbTe, which also has a rocksalt structure, is not a TCI and its surface states are gapped. Notwithstanding, a topological phase transition can be induced either by pressure or alloying. In particular, as anticipated above, Pb$_{1-x}$Sn$_x$Te is a topologically non-trivial material ($n_M=-2$) for $x\gtrapprox 1/3$~\cite{xu2012observation,Dziawa2012}. Actually, a band inversion as a function of the alloy composition was reported long time ago~\cite{PhysRevLett.16.1193}.

Thin films of these TCIs also show nontrivial topological behavior. However, hybridization between the surface states opens a gap that decreases for increasing layer thickness~\cite{Ozawa2014SnTe001,Liu2013thickness}. In fact, a non-monotonic damped oscillation has been found, that it is ultimately related to the alternation between nonsymmorphic and symmorphic symmetry with the number of atomic layers~\cite{Ozawa2014SnTe001,Araujo2018SnTe001}. Specifically, if a $[001]$-oriented SnTe slab has an even number of layers, it possesses nonsymmorphic symmetry. This reduces the hybridization between surface states, whereas for an odd number of layers the symmetry is symmorphic, and more sizeable band gaps appear. 

Additionally, symmetry-protected surface states in TCIs can be tuned by introducing impurities. The effects of doping on the electronic structure of bulk SnTe have been addressed in Refs.~\citenum{Hoang2010, Zhao2018}. As for the $(001)$ surface, In doping~\cite{Schmidt2020}, hydrogen adsorption \cite{D1NR05089C} and diluted impurities~\cite{PCCP21} have been considered. Doped SnTe, specially by In impurities, is an ongoing subject of discussion regarding topological superconductivity \cite{superconductor2}.

Impurities in TCI thin films may open a gap or modify the surface states, allowing for the control of topological phase transitions in these materials and constituting an additional tool to modulate their properties~\cite{PhysRevLett.112.046801,PCCP22}. Indeed, tuning the electronic properties by doping or by applied external fields opens a way for the application of these materials in nanoscale devices \cite{Liu2013thickness}. Such modification by doping may be also relevant for thermoelectric applications, another arena in which topological materials have been proven to be of interest \cite{Hoang2010, Zhao2018}. 

In this work, the effect of non-magnetic impurities in the topological crystalline insulator SnTe is studied. Specifically, we consider Sn substitution by a Sb atom, which acts as a donor impurity.  In fact, Sb-doped SnTe has been found to be a superconductor~\cite{superconductor}. Here, we focus on the changes in the TCI dispersion relations induced by the presence of the Sb impurity. As mentioned above, doping may alter or modify the surface states by hybridization or just by opening a gap, therefore changing the electronic properties of the system. This is the main issue we address herein. We explore both semi-infinite and slab geometries, with a twofold goal. Firstly, ultrathin layers of TCIs are currently grown and incorporated in quasi-2D materials and devices as building blocks or integrating parts, so the thin slab geometry has an interest in itself. Secondly, in most first-principles calculations, a finite slab geometry is employed to model surfaces, so it is equally important to evaluate the differences between finite-size and semi-infinite calculations aimed at modeling the same system. To this end, DFT-based \textit{ab-initio} simulations are used to characterize SnTe and the effect of the non-magnetic donor impurity near the crystal surface. We analyze the hybridization between surface and impurity states as well as the effect of the impurity position on the electronic structure and crystal symmetries. Furthermore, both perturbation theory and continuum models are proposed. The former is used to predict the behavior of the Dirac cone upon doping, while the latter describes the impurity states.

Our main results are summarized as follows: 
(i) The Dirac cones appearing at the TCI surfaces are shifted down upon the $n$-doping due to Sb substitutional impurities. 
(ii) The shift of the Dirac cones is strongly dependent on the position of the impurity with respect to the surface. 
(iii) The width of the impurity band presents an even-odd behavior with the layer position of the dopant, both in the semi-infinite and the slab geometries; we relate it to the location of the nodes of the wave function with respect to the surface. 
(iv) In the slab system, a gap opens in the Dirac points due to coupling of the states at opposite surfaces, even though they are almost 50 \AA\ apart. We verify that the surface states remain gapless in the semi-infinite geometry, with unaltered spin textures. 
(v) Finally, we derive continuum models in order to extrapolate the behavior of this doped system for larger sizes, unreachable with more accurate, \textit{ab-initio} simulations.

Our work provides a complete picture of the interplay between impurity and surface states in TCIs, highlighting the differences between finite-sized slabs and single surfaces, and additionally including simpler continuum models which yield a better interpretation of the numerical first-principles results. 

\section{First-principles approach}
\subsection{Geometry and methods}

Our calculations were performed using the pseudopotential DFT SIESTA code \cite{Soler2002,siesta2020} within the generalized gradient approximation under the PBE parametrization (GGA-PBE) for the exchange and correlation functional~\cite{PBE1996}. Spin-orbit coupling was considered within the fully-relativistic pseudopotential formalism~\cite{Cuadrado_2012} for all calculations unless otherwise stated. Accordingly, a relativistic pseudopotential was employed, following the Troullier-Martins method \cite{PhysRevB.43.1993}.

A double-$\zeta$ singly-polarized basis set was chosen for all the atoms involved, with a radial extension of the strictly localized orbitals determined from a confinement energy (energy shift) of \SI{100}{\milli\electronvolt}. The resolution of the real space grids was set to around \SI{0.06}{\angstrom\cubed} (equivalent to a mesh cutoff of \SI{600}{\rydberg}). For the integration over the BZ we employed $k$-supercells of size around (17$\times$17) relative to the SnTe-(1$\times$1) bulk unit cell, with a temperature in the Fermi-Dirac occupation function of \SI{100}{\milli\electronvolt}.

A lattice parameter of $\SI{6.39}{\angstrom}$ was used following Refs.~\citenum{Hsieh2012SnTe,Hoang2010}, which yields a satisfactory agreement to experimental results and a successful magnitude of the band gap~\cite{PhysRevLett.16.1193}. 
This allows us to model rocksalt room-temperature SnTe, avoiding the aforementioned rhombohedral phase transition \cite{xu2012observation}. In order to tackle the robustness of the surface states against film thickness, several slab calculations were performed using different number of layers (not shown here). We found an optimal compromise between accuracy and our computational capability for a nonsymmorphic 16 atomic-layer SnTe slab. Therefore, all the slab calculations presented in this work have been performed with this thickness.

Substitutional donor impurities have been considered by replacing a Sn atom by Sb within a c$(4\times4)$ supercell ---that is, in a defective layer one out of eight Sn atoms is substituted by a Sb atom. We have considered up to four different locations of the Sb: at the top surface layer (referred to as Sb1 hereafter), the second layer (Sb2), the third layer (Sb3) and the fourth layer (Sb4). Furthermore, in order to avoid net dipoles in the slab, a second Sb substitutional impurity was always placed at an equivalent position, but with respect to the bottom layer [see Figure~\ref{fig1}(a)]. In this way, both impurities are located at the same distance from the corresponding nearest surface. Thus, the simulated impurity doping in the slabs corresponds to a small value of $x = 0.016$.

Atomic relaxation was performed for each structure. For the Sb1 slab, the two top and bottom layers were relaxed while for the Sb2, Sb3 and Sb4 slabs, the $3$, $4$ and $5$ layers counted from the top and bottom surfaces were included in the relaxations until forces on the atoms were below \SI{0.02}{\electronvolt/\angstrom}. The top and bottom layers of the final equilibrium structure are the most affected by the relaxation. To be specific, some atoms of the Sb1 and Sb3 slabs are displaced up to \SI{0.17}{\angstrom} compared to \SI{0.14}{\angstrom} for the Sb2 and Sb4 slabs.

Although minimal, the hybridization of opposite surface states still has an effect on the band structure. Hence, the electronic and spin structure of the Sb-doped SnTe(001) surface will be mainly presented as calculated under a semi-infinite geometry \textit{via} projected density of states and magnetization maps, $\mathrm{PDOS}({\bm k},E)$ and $M_\alpha({\bm k},E)$ respectively, with $\alpha=x,y,z$ indicating the direction~\cite{Green1,Green2}. To this end, we have employed surface Green's function matching techniques in order to couple the Hamiltonian matrix elements of the top half of the slab with those of a bulk SnTe calculation. Self-consistency was preserved in the matching process, as reflected by the fact that maximum deviations were always smaller than \SI{10}{\milli\electronvolt} between the on-site energies of the Sn and Te central layers of the slab and those in the SnTe bulk. $\mathrm{PDOS}({\bm k},E)$ and $M_\alpha({\bm k},E)$ maps were computed with a resolution of \SI{0.003}{\angstrom^{-1}} in reciprocal space and \SI{5}{\milli\electronvolt} in energy. Accordingly, the self-energy that determines the spectral width of any surface states (imaginary part of the energy entering the Green's function) was set to \SI{5}{\milli\electronvolt}. For zoom-ins of the Dirac points we computed additional maps with an increased resolution (\SI{0.0016}{\angstrom^{-1}} and \SI{3}{\milli\electronvolt}) around the $\bar{\Gamma}$-point and over a small energy window. In the semi-infinite calculations, a single impurity is used. For topologically-protected surface states, we expect them to remain unaltered, since the placement of a single impurity does not change the $(110)$ mirror symmetry, as outlined in Figure \ref{fig1}(c). On the contrary, the 16 atomic layer slabs do break this symmetry when taking into account two impurities, as explained before. Since two-dimensional (few-layer) topological crystalline insulators are also the subject of great attention, we consider this comparison to be relevant. 

\subsection{Ab-initio results}

\begin{figure}[h!]
    \centering
    \includegraphics[width=0.48\textwidth]{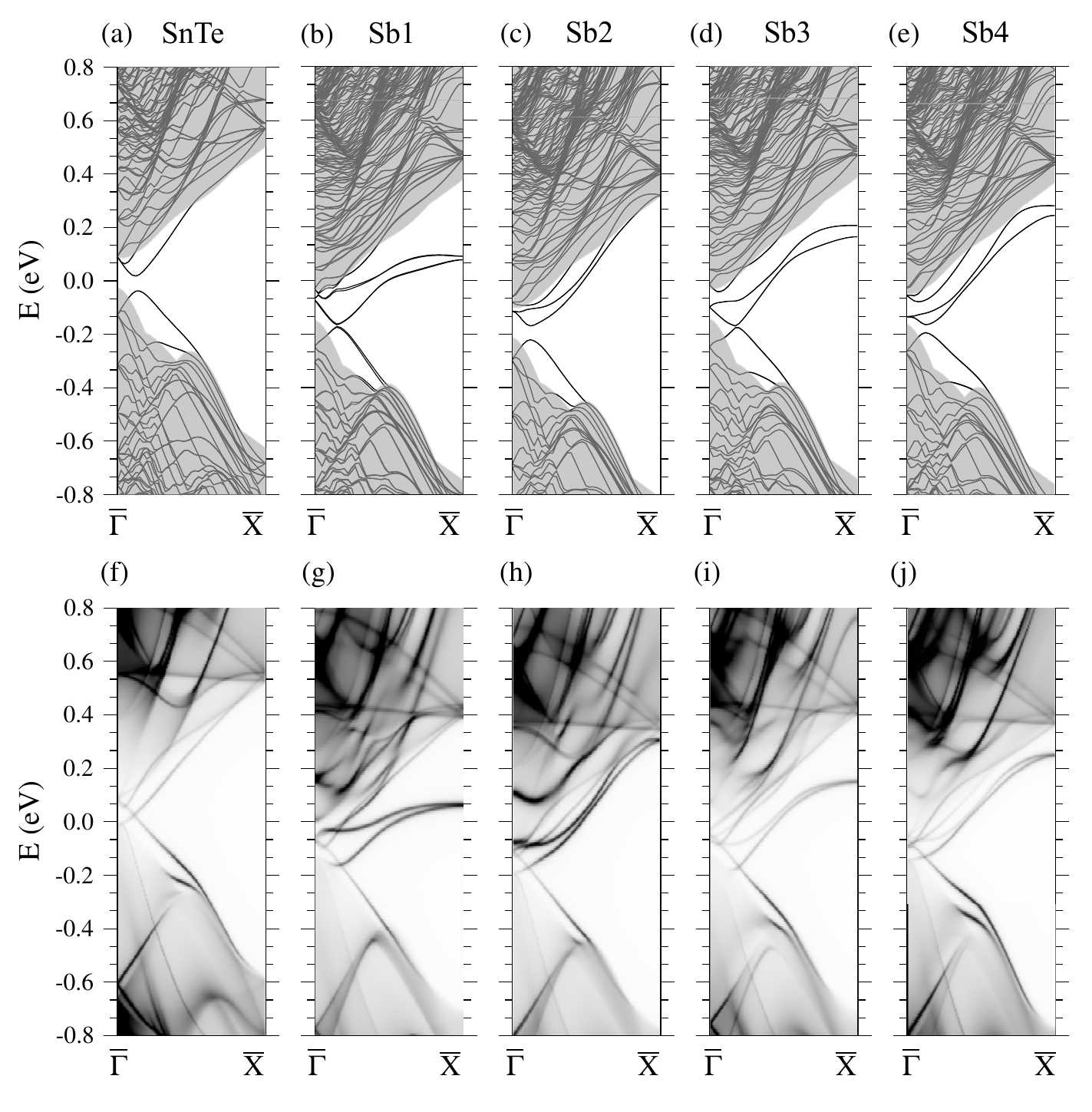}
    \caption{Electronic structure of pristine SnTe (left column) and Sb-doped SnTe, with the donor impurity located in the first,
    second, third and fourth atomic layer. Top panels shows the band structure of the 16-layer slabs, as well as the bulk band projections of pristine SnTe represented by the gray shaded area. Bottom panel displays the projected density of states $\mathrm{PDOS}({\bm k},E)$ on the first atomic layers for the semi-infinite system.}
    \label{fig2}
\end{figure}

\begin{figure}[t!]
    \centering
    \includegraphics[width=0.49\textwidth,trim={0cm 0.0cm 0.0cm 0.0cm},clip]{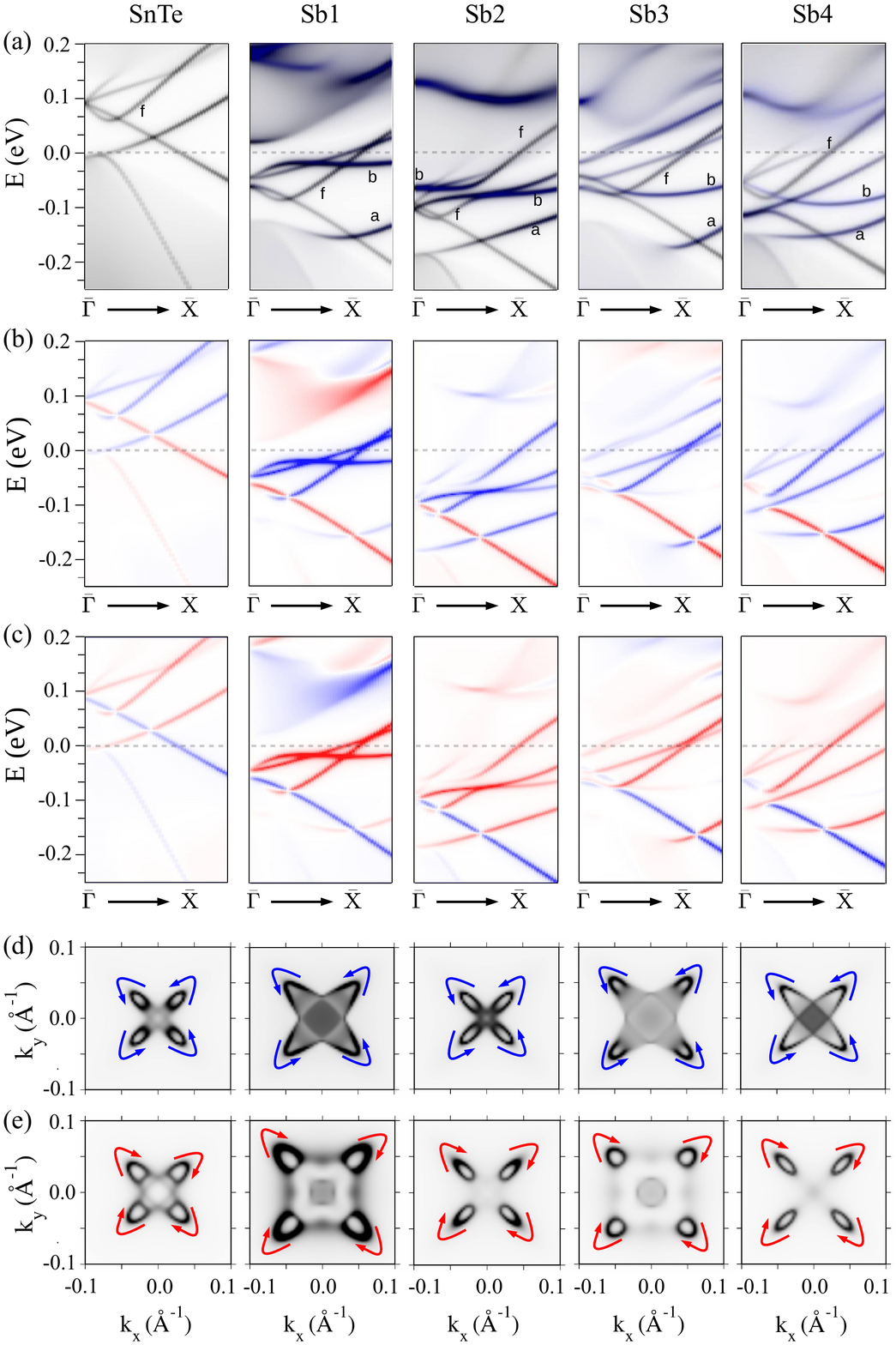}
    \caption{Zoomed (a)~$\mathrm{PDOS}({\bm k},E)$ as well as (b)~$M_{x}$ and (c)~$M_{y}$ maps of the semi-infinite structure. In panel (a), the Sb impurity contribution is highlighted in blue. $\mathrm{PDOS}(k_x,k_y,E)$ maps at an energy $\SI{20}{\milli\electronvolt}$ (d)~below and (e)~above the Dirac point. The spin textures are depicted with arrows that outline the spin direction. See main text for an explanation of labels $a$, $b$ and $f$ in the upper panel.}
    \label{fig3}
\end{figure}

Top panel of Figure \ref{fig2} displays the effect of the substitutional Sb on the surface electronic structure for the four locations of the impurity as it is moved from the surface layer towards the bulk, namely Sb1 to Sb4. The band structures have been computed along the $\bar{\Gamma}-\bar{X}$ direction for 16-layer slabs (that is, under an ultrathin film geometry). Note that, due to band folding, the $\bar{X}$ point in the (1$\times$1) BZ back folds into the $\bar{\Gamma}$ point of the supercell's BZ. For the sake of comparison, we also include in the left column the ${\bm k}$-folded SnTe pristine surface computed under the same c$(4\times4)$  supercell as that used for the defective structures. There are several features immediately apparent in the plots. First of all, in all defective surfaces the Sb develops a split impurity band that disperses across the gap. Interestingly, the band is narrower when the impurity is located at an odd-numbered layer (particularly the Sb1 case but also Sb3), due to a higher localization of the states at the surface. On the contrary, when the impurity is located at an even-numbered layer (Sb2 and Sb4 cases), the band becomes more dispersive and runs closer to the edge of the bulk conduction band. Second, the presence of the impurity induces a clear $n$-type doping as the Fermi level $E_F$ shifts towards the minimum of the conduction band in all defective structures, while it is pinned just above the valence band (VB) maximum for the pristine surface. This doping effect will be further addressed in Sec.~\ref{sec:cont} using a minimal continuum model. Despite the large thickness of the slabs employed, all Dirac points present a gap due to the long decay length of the topological surface states, so that there is still a sizeable interaction between the top and bottom surfaces. The size of this gap is minimal for the odd cases (Sb1 and Sb3), being around $5\,$meV, while it attains substantially larger values for the even cases. Note that it is around $50\,$meV for Sb2 and $30\,$meV for Sb4.  For the pristine slab, for which the gap is due to finite-size effects, it is $55\,$meV. Moreover, the breaking of the $(110)$ mirror symmetry in the two-impurity geometry used in the slabs affects the symmetry-protected surface states and contributes to the gap opening, as depicted in panels (b)--(e). Although such even-odd behavior in the Dirac cone gap value with the impurity position is reminiscent of the nonsymmorphic-symmorphic effect with respect to the number of layers previously reported~\cite{Ozawa2014SnTe001,Araujo2018SnTe001}, we note that in our case its origin should be different, because all slabs are composed of the same number of layers. 

The bottom panel of Figure~\ref{fig2} displays the equivalent electronic structures to the top panel but in the form of $\mathrm{PDOS}({\bm k},E)$ maps, since they have been computed under a semi-infinite geometry (see previous section). The Sb-doped surfaces, overall, show a similar behavior to that found for the 16-layer slabs, namely, the appearance of Sb impurity bands across the gap showing an even-odd effect in their dispersion, a clear $n$-type doping and the presence of Dirac cones slightly away from $\bar{\Gamma}$. However, these Dirac cones remarkably persist in all defective cases of the semi-infinite structure. The splitting of the impurity bands induced by the spin-orbit coupling remains small until it approaches the $\bar{\Gamma}$ point, where the two branches start to deviate from each other by more than \SI{100}{\milli\electronvolt}.

Focusing now on the electronic structure around the $\bar{\Gamma}$ point, the most relevant feature is the survival of a gapless Dirac point in all structures, which is consistent with the fact that the $(110)$ mirror plane still holds in the doped structures and, hence, the Dirac points remain symmetry-protected. However, there are qualitative differences with respect to the pristine surface which are best seen in figure~\ref{fig3}, where zoom-ins with an enhanced resolution are presented. First of all, and in accordance with the $n$-type doping mentioned above, the Dirac points consistently shift by around 150$-$200~meV towards lower energies. Precisely at the $\bar{\Gamma}$ point the interpretation of the electronic structure becomes more complex due to the profusion of extra bands arising from both the Sb impurity and the band folding. 

All relevant bands within the gap are better resolved in figure~\ref{fig3}(a), where we present enlarged views of the $\mathrm{PDOS}({\bm k},E)$ maps. The projections have been taken over the surface SnTe layer, superimposing the Sb contributions in blue. 

For the clean surface case, the branches forming the Dirac cone can be clearly identified, as well as the Sb-split impurity bands, denoted as $a$ and $b$. There is a further band inside the gap, labeled by $f$, corresponding to the back-folded topological surface state along $\bar{X}-\bar{M}$ [see figure~1(d)]. 
Since the supercell size is much smaller than the effective Bohr radius, impurities cannot be regarded as truly isolated and the impurity bands become dispersive. In addition, further broadening of the bands arises from the hybridization of impurity states with Bloch states of the conduction band of SnTe. In fact, the impurity states are mainly composed of Sb and Sn contributions, as depicted in the Appendix A for both slab and bulk band structures. Band inversion between Sn and Te orbitals also persists in the doped material, as discussed and shown in Appendix A.

In the lower panels (b) and (c) of figure \ref{fig3}, magnetization maps along the $x$- and $y$-axis, $M_{x}({\bm k},E)$ and $M_{y}({\bm k},E)$, respectively, are shown. We omit the out-of-plane component $M_z({\bm k},E)$ since it is always negligible except for Sb1. As it could be expected, the in-plane spin orientations of the surface band remain unaltered with respect to the pristine case. Indeed, this is best seen in the PDOS($k_x,k_y$) maps shown in the lower panels, computed at energies \SI{20}{\milli\electronvolt} below [panel (d)] and above [panel (e)] the location of the Dirac point. 

\begin{figure}[h!]
    \centering
    \includegraphics[width=0.47\textwidth]{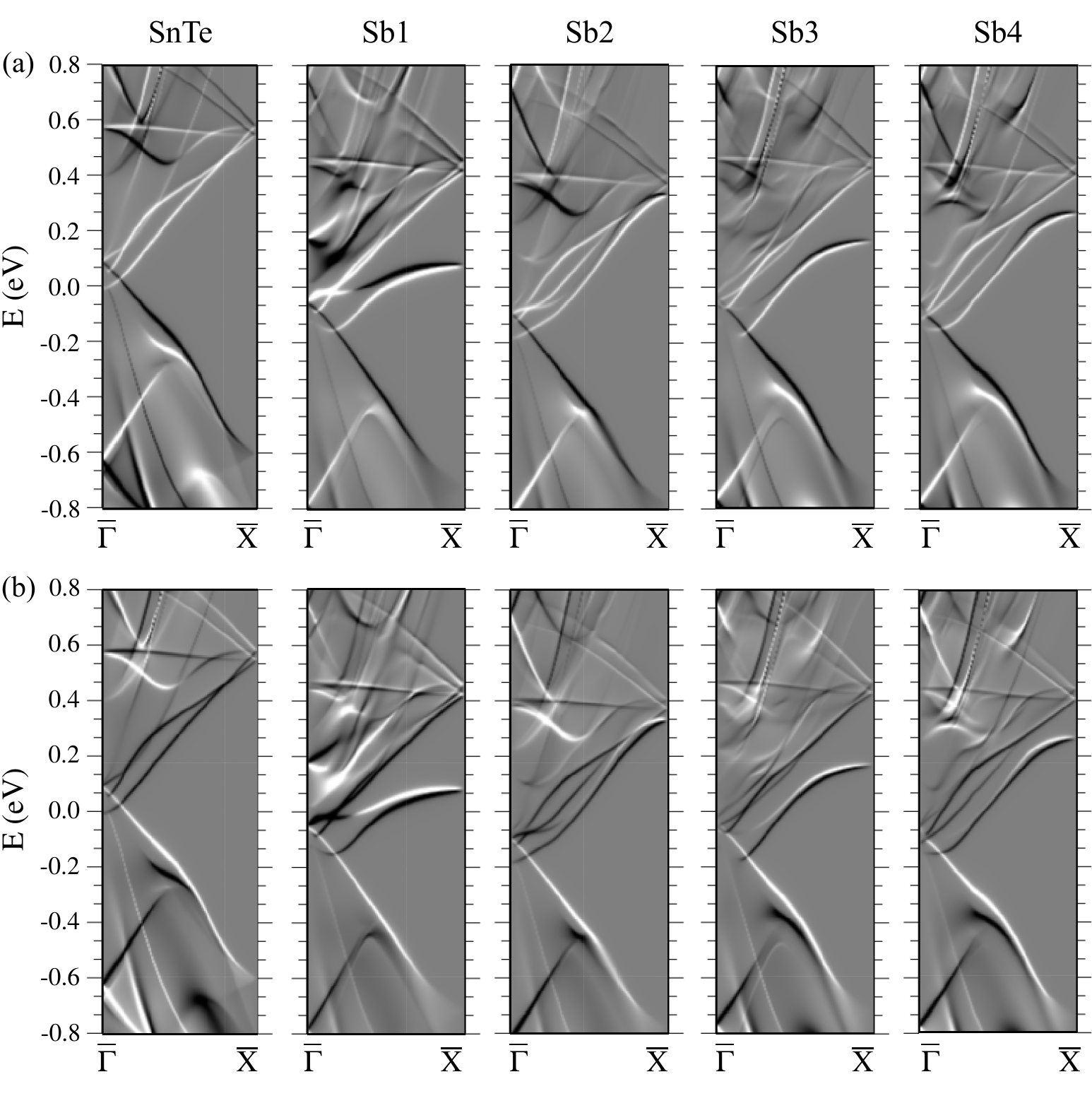}
    \caption{Magnetization maps (a)~$M_x({\bm k},E)$ and (b)~$M_y({\bm k},E)$ for the semi-infinite pristine and Sb-doped structure, projected on the first atomic layers. White and black shades account for positive and negative values of the magnetization $M_\alpha({\bm k},E)$, respectively.}
    \label{fig4}
\end{figure}

The spin texture for all structures is shown in figure~\ref{fig4} in the form of $M_{x/y}(k,E)$ maps. Note that pristine SnTe only shows in-plane spin polarization due to time-reversal and mirror symmetries \cite{Wang2013SnTe001}. The effect of folding on the spin texture of pristine SnTe is explicitly detailed in the Appendix A, showing additional $M_{x/y}(k_x,k_y)$ maps near the Dirac point for all structures. A close inspection of the spin textures reveals that the spin orientation of the topological surface states always remains unchanged with respect to the pristine surface. More surprising is the behavior of the Sb bands: close to $\bar{X}$ they behave as standard Rashba-split states, with a reduced separation in energy between $a$ and $b$ of only 10$-$20~meV while they present opposite magnetizations. In fact, this is the only region where out-of-plane magnetization, $M_z({\bm k},E)$, is not negligible. However, as
they disperse towards $\bar{\Gamma}$, band $a$ keeps its spin orientation but
$b$ undergoes an abrupt spin inversion and ends up with the same orientation as $a$. Note that the spin texture of the impurity states is not as robust as that of the Dirac cone, since it varies along the high-symmetry path and is dependent on the choice of atom projections.

\begin{figure}[h!]
    \includegraphics[width=0.49\textwidth]{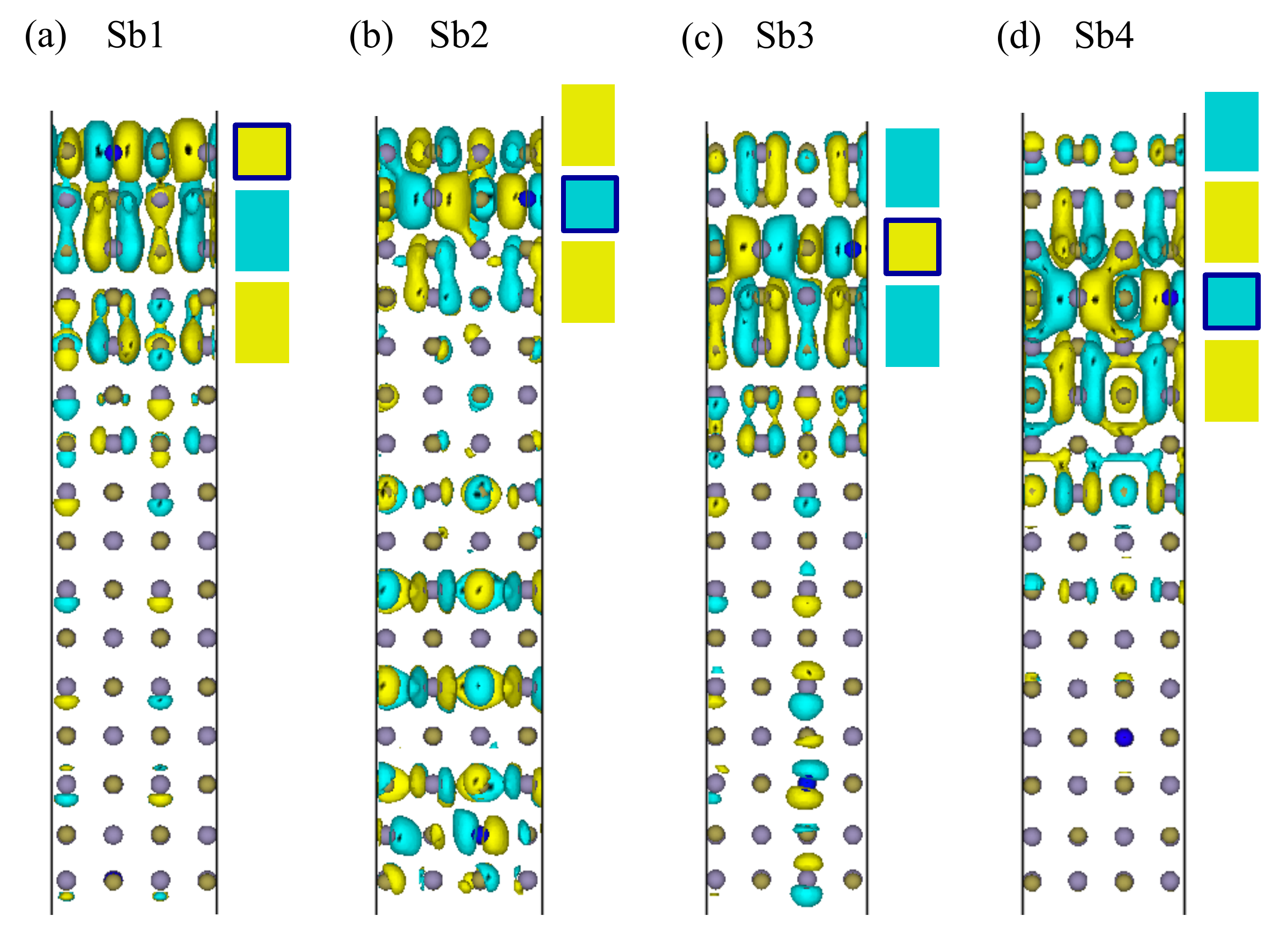}
    \caption{Wave functions for one of the impurity bands at $\bar{\Gamma}$. Yellow and blue colors represent positive and negative values, respectively. The colored rectangles next to each panel, which are two atomic layers wide, outline the alternation of regions with positive and negative values of the wave function in the $z$-direction. The layer where the impurity is located is highlighted in dark blue.}
   \label{fig5}  
\end{figure}

Given such ultrathin layers, the energy shift of the Dirac point can be related to quantum size effects. In order to clarify this point, we study the spatial distribution of the wave function for the impurity band, depicted in figure~\ref{fig5}. A series of nodes appear right after the impurity and approximately each two atomic layers in the $z$-direction. When the impurity is located at an odd atomic layer, there is an even number of layers above it, so the wave function 
amplitude is almost zero at the surface layer. In contrast, when the impurity is placed in an even layer, there is an odd number of layers above it and the wave function is nonzero at the surface, therefore favoring hybridization with the surface states and a larger band gap. Since the derivative of the wave function is related to the kinetic energy operator, it is also expected for these impurity bands to move upwards, as shown in both slab and semi-infinite calculations in figure~\ref{fig2}.
Thus, we come to the conclusion that this energy shift can be also seen as a size or quantum confinement effect. 

\section{Continuum model}
\label{sec:cont}

In order to have an insight into the effect of the impurity upon the Dirac cones and the behavior of the impurity band when the position of the donor Sb atom is varied, we resort to a simple continuum model of a TI. 
The SnTe family has two non-equivalent valleys, $L_1$ and $L_2$, in the Brillouin zone. However, a single valley approach successfully explains the existence of robust surface states on crystal surfaces such as $(001)$, $(110)$ or $(111)$~\cite{Hsieh2012SnTe}. Hence, in this work we focus on states close to one of the $L$ points of the Brillouin zone and neglect other valleys since surface states are stable against gap opening by valley mixing.
It consists of an isotropic two-band spin-full continuum Hamiltonian in a particle-hole symmetric scenario.  In fact, this model has been already employed for TIs~\cite{shen2017book,Liu2010BiSe3}. The Hamiltonian in momentum space is given by (we set $\hslash=1$ hereafter)
\begin{equation}
    \ham_0 = \epsilon \mathbb{1}_4+ v \, \vec{\alpha}\cdot \vec{k} + \left( m - B \vec{k}^2\right) \beta~ ,
    \label{model-Hamiltonian}
\end{equation}
where $\epsilon$ is a constant energy term, $\mathbb{1}_4$ is the $4\times 4$ identity matrix, $v$ is the Fermi velocity, $m$ is the mass parameter related to the gap and the constant $B$ takes into account the quadratic terms in momentum. In the former expression, $\vec{\alpha}$ is the vector of Dirac matrices and the momentum-vector defined as $\vec{\alpha} = (\alpha_x, \alpha_y, \alpha_z)$. We choose the Dirac matrices expressed in terms of the Pauli matrices in the following basis: $\alpha_i \equiv \sigma_x \otimes \sigma_i~$, $\beta \equiv \sigma_z \otimes \sigma_0~$. This simple model has the minimum elements needed to show topologically protected surface states if $mB>0$ and the translational symmetry is broken by making a finite-size system~\cite{Shen2010TopologicalEquation}.

Modifications of this model, including anisotropy and on-site momentum dependent terms, have been successfully proposed to study the \ce{Bi_2Se_3} family of materials~\cite{Zhang2009Bi2Se3,Liu2010BiSe3}. In fact, the model has also been applied to describe the $(001)$ surface states of \ce{SnTe} including just the leading terms linear in momentum~\cite{Zhang2012Bi2Se3_SnTe,Hsieh2012SnTe,Liu2013SnTe,Ozawa2014SnTe001,Araujo2018SnTe001,Wang2013SnTe001}. Since our first goal within the continuum approach is to analyze the effect of an impurity in close proximity of an isolated surface, quadratic terms are indispensable to have a nontrivial topology in a vacuum-\ce{SnTe} interface~\cite{shen2017book}. We expect to find the same shifting behavior as predicted by Araújo \textit{et at.}~\cite{Araujo2018SnTe001} and our DFT calculations. Thus, we perform first-order perturbation theory (PT) on the surface states obtained within this formalism with the aim of elucidating the effect of the impurity on the gapless surface states. Secondly, in order to describe the energy dependence of the impurity states with respect to the impurity position, we solve a simplified version of the Hamiltonian presented in equation~\eqref{model-Hamiltonian}, summing up all terms in the perturbation series using a Green's function method. These two approaches allow us to investigate complementary aspects of the problem.

\subsection{Perturbation theory approach}

With the aim of comparing PT results with DFT calculations, we consider the case of a semi-infinite system which extends from $z>0$, being infinite in the other perpendicular directions. The surface states are derived in Appendix B 
imposing Dirichlet boundary conditions at the plane $z=0$. In this way we obtain two Kramers partners that only differ in the spinor part $\Phi^0_\pm$
\begin{equation} \label{PT:eq:surfStates}
        \psi_\pm = A_s (e^{-\la_1 z} - e^{-\la_2 z}) \exp{[i(k_x x + k_y y)]} \Phi^0_\pm~.
\end{equation}
In the former expression, the inverses of the decay lengths are given by $\la_1$ and $\la_2$, which are functions of the in-plane momenta and $A_s$ is a normalization factor (see Appendix B). The dispersion of the states is linear in the in-plane momenta, yielding the Dirac cones of the surface states inside the bulk gap $E_\pm = \epsilon  \pm v k_\parallel \sign(B)~$, where $k_\parallel= \sqrt{k_x^2+ k_y^2}$~.

For the purpose of obtaining the corrections on the energy, we perform first-order PT, modeling the impurity with a Coulomb-like interaction
\begin{equation} 
    \Delta \mathcal{H}=  - \frac{2 R_y^* a_B^*}{\sqrt{(z-z_0)^2+x^2+y^2}} ~,
    \label{PT:eq:DeltaH}
\end{equation}
where $z_0$ is the depth where the impurity is located, and $R_y^*$ and $a_B^*$ are the effective Rydberg energy and Bohr radius in \ce{SnTe}~\cite{knox1963Book,ibach2009Book}. The former constants provide the correct units to the expression and scale the interaction due to the medium, as detailed in Appendix B.

The first-order correction in PT is given by 
\begin{equation} \label{PT:eq:PertEn}
    \Delta E^{(1)}_\pm = \bra{\psi_\pm} \Delta \ham_0 \ket{\psi_\pm} ~,
\end{equation} 
where $\psi_\pm$ are the surface states as defined in equation~\eqref{PT:eq:surfStates} for the impurity-free case. Due to the trivial structure of the correction in the spin space, both states, $\psi_+$ and $\psi_-$, show the same correction. Therefore, the cones are just shifted by the same amount and no gap opens.

\begin{figure}[htb]
    \centering
    \includegraphics[width=0.5\textwidth]{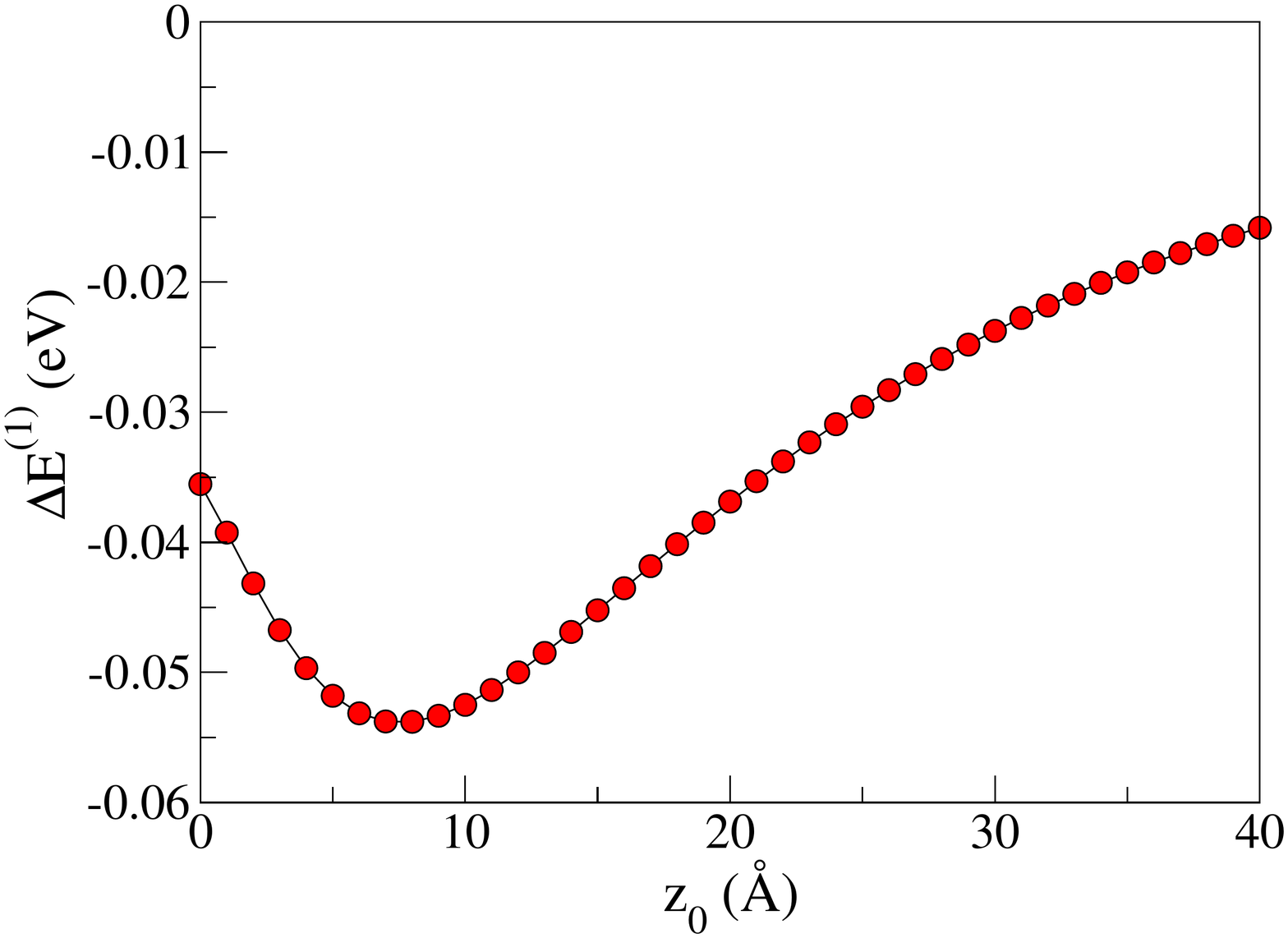}
    \caption{Displacement of the Dirac point $\Delta E ^{(1)}$ as a function of the impurity depth $z_0$, obtained from PT calculations.} 
    \label{fig:PTResults}
\end{figure}

Figure~\ref{fig:PTResults} shows the displacement of the Dirac point as a function of the impurity depth $z_0$. Similarly to Figure~\ref{fig2}, a shift of the Dirac point as the impurity is placed farther from the surface is observed, in opposition to hole doping~\cite{Schmidt2020}. While DFT calculations are difficult to address due to their computational cost when the impurity is far from the surface, this minimal model allows us to predict a return to the pristine case for large $z_0$. Despite the aforementioned limitations, the model describes qualitatively well the phenomenon and it can be used as a simple approach to the effect of the impurity on the surface bands.

\subsection{Green's function approach}\label{sec:GFA}

In the previous subsection, the interaction of the electron with the impurity was assumed weak so that the first-order perturbation theory was a reasonable approximation. We can go beyond this limit, summing up all terms of the perturbation series by means of the Green's function approach. To this end, we start with equation~(\ref{model-Hamiltonian}) and neglect the quadratic in momentum term ($B=0$). The linear approximation (Dirac-like equation) is unable to explain the existence of surface states in a TI unless a normal semiconductor is attached, thus forming a heterojunction. In this way, there exists a band inversion along the direction normal to the surface and interface states can appear (see, e.g., Ref.~\cite{shen2017book} and references therein). 

The envelope functions of the electron states near the band extrema are then determined from the following Dirac-like Hamiltonian 
\begin{equation}
\mathcal{H}_0=v\,{\bm\alpha}\cdot{\bm k}+m(z)\,\beta\ ,
\label{eq:GFA-01}
\end{equation}
where we assume aligned and same-sized gaps for simplicity. The origin of energy is set at the center of the gaps ($\epsilon=0$) and the mass term is given as $m(z)=m\sign(z)$. The Hamiltonian~(\ref{eq:GFA-01}) acts upon the envelope function ${\bm\chi}({\bm r})$, which is a four-component vector composed of the two-component spinors ${\bm\chi}_{+}({\bm r})$ and ${\bm\chi}_{-}({\bm r})$. It decays as $\exp(-|z|/d)$ along the direction normal to the surface, with $d=v/m$, and the interface dispersion relation is a single Dirac cone $E({\bm k}_{\parallel})=\pm  v|{\bm k}_{\parallel}|$.

The presence of the impurity, located at ${\bm r}_0=(0,0,z_0)$ without loss of generality, breaks the translational symmetry in the surface and the in-plane momentum is no longer conserved. Expressing distance in units of $d$ and energy in units of $m$, the Dirac-like equation for the envelope function can be cast in the form
\begin{multline}
\Big[ E\mathbb{1}_4+i\,\alpha_z\,\frac{\partial\phantom{z}}{\partial z}+i\,{\bm \alpha}_\parallel\cdot \nabla_\parallel-\beta \sign(z) \\ -V({\bm r}-{\bm r}_0)\mathbb{1}_4 \Big] {\bm\chi}({\bm r})={\bm 0}\, .
\label{eq:GFA-02}
\end{multline}
The actual interaction potential $V({\bm r}-{\bm r}_0)$ will be replaced by a non-local separable pseudo-potential of the form
\begin{equation}
V({\bm r}-{\bm r}_0) {\bm\chi}({\bm r})\rightarrow
\lambda \omega({\bm r}-{\bm r}_0) \!\int\! \mathrm{d}^3{\bm r}^{\prime} \omega({\bm r}^{\prime}-{\bm r}_0){\bm\chi}({\bm r}^{\prime})\ ,
\label{eq:GFA-03}
\end{equation}
where $\lambda$ is a coupling constant and $\omega({\bm r}-{\bm r}_0)$ is a real function referred to as shape function hereafter. It is worth mentioning that this replacement is exact, in the sense that it is always possible to find a non-local separable pseudo-potential (or a sum of them) able to reproduce any set of given electronic states and, consequently, there is no theoretical limitation to the numerical accuracy with which physical results can be obtained~\cite{Glasser77}. In spite of its seemingly more complicated form, equation~(\ref{eq:GFA-03}) is amenable to analytical solution for any arbitrary shape function~\cite{Knight63,Sievert73,Lopez02,Adame91,Adame95,Gonzalez_Santander13}.

The Green's function for the unperturbed problem ($\lambda=0$) and the Green's function for the total Hamiltonian in terms of the transition operator are derived in Appendix B
. Poles in the complex plane of the transition matrix, $E=E_\mathrm{r}-i\Gamma$, where $E_\mathrm{r}$ and $\Gamma$ are real magnitudes, yield the energies of the impurity states. Resonances within the gap are given by the condition $|E_\mathrm{r}|<1$. According to the model presented, an impurity embedded in a \emph{bulk\/} semiconductor can support a truly bound state. Lengthy but straightforward calculations lead to the following energy for the bound state 
\begin{equation}
E_\mathrm{b}=\sign(\lambda)\left(1-\frac{4\pi}{|\lambda|k_\mathrm{c}}\right)\ ,
\label{eq:GFA-11}
\end{equation}
in the limit $\omega({\bm r}-{\bm r}_0)\to \delta({\bm r}-{\bm r}_0)$ and $k_c$ is a cut-off momentum that regularizes the $\delta$ function. Since only states within the gap correspond to bound states in this model ($|E_\mathrm{b}|<1$), there exists a minimum value of the product $|\lambda|k_\mathrm{c}$ to induce bound states. If $|\lambda|k_\mathrm{c}<2\pi$ the impurity cannot bind electrons ($\lambda<0$) or holes ($\lambda>0$) in bulk semiconductors. 

\begin{figure}[ht!]
    \centering
    \includegraphics[width=0.4\textwidth]{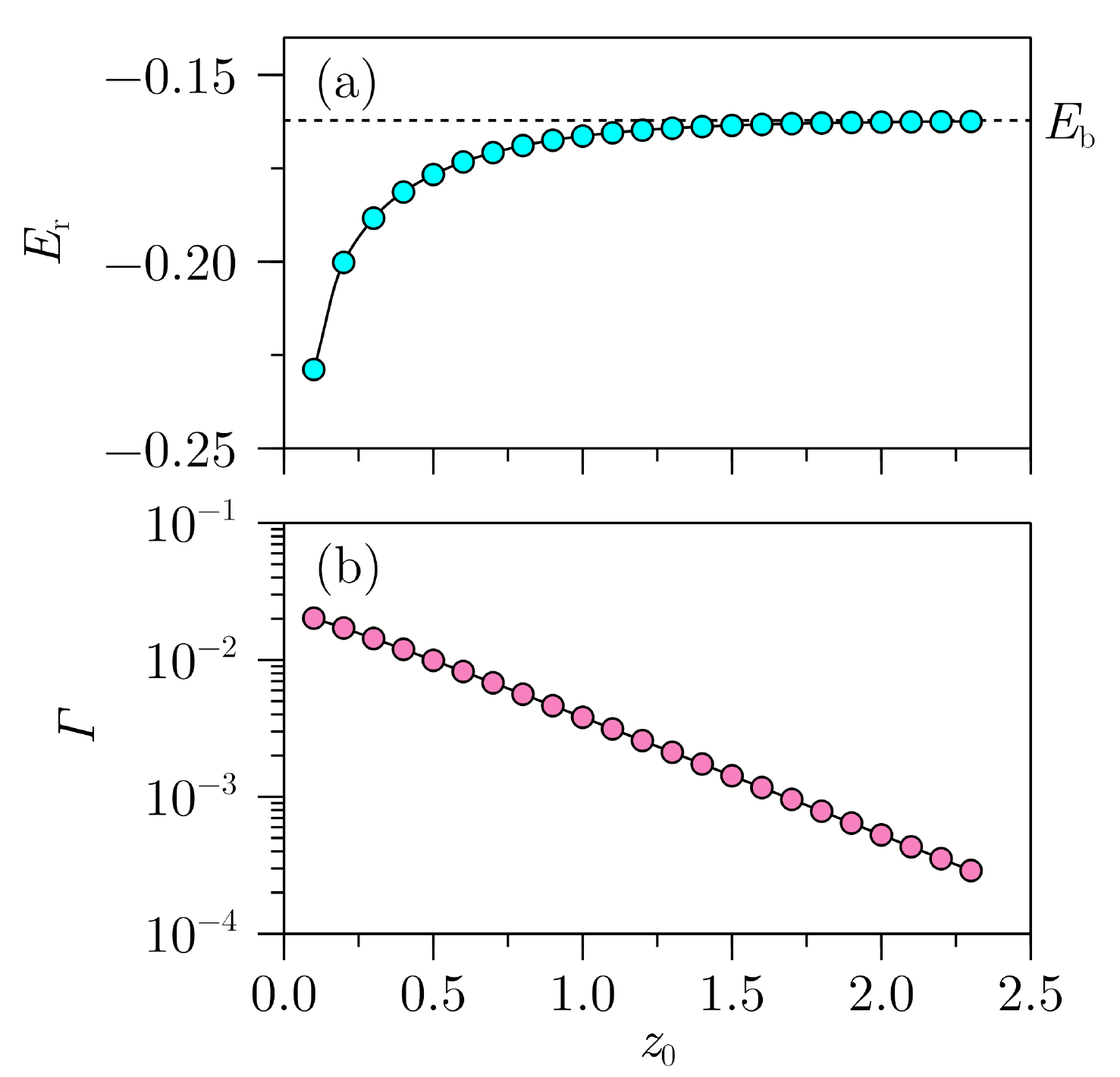}
    \caption{Real and imaginary parts of the complex poles $E=E_\mathrm{r}-i\Gamma$ of the transition matrix as a function of the impurity position. The parameters of the interaction potential are $\lambda=-0.15$ and $k_\mathrm{c}=100$. Dashed line indicates the energy of the impurity bound state in the bulk semiconductor [see equation~(\ref{eq:GFA-11})].}
    \label{fig:Evsz0}
\end{figure}
    
Figure~\ref{fig:Evsz0} shows the real and imaginary parts of the poles of the transition matrix as a function of the position of the impurity close to the     \emph{heterojunction}, when $k_\mathrm{c}=100$ and $\lambda=-0.15$. The real part $E_\mathrm{r}$ approaches the value in the bulk material $E_\mathrm{b}$ when increasing $z_0$, as expected. These impurity states move to higher energies as the distance from the surface increases, as in Ref.~\citenum{PhysRevB.50.17074}. The alternating behavior found in DFT calculations, with different energy shifts of the impurity bands for even (odd) slabs, is not reflected in this model since simpler, exponentially-decaying wave functions are used here.

\section{Conclusions}\label{end}

Doping a SnTe slab with a non-magnetic Sb impurity breaks mirror symmetry, which affects the well-known topologically-protected surface states of pristine SnTe. Additionally, impurity doping has an effect on the band structure: it creates impurity bands which hybridize with the aforementioned surface states. We have studied this doped system by resorting to first-principles calculations and continuum models. DFT calculations have been performed to study a finite structure, i.e., a slab with two surfaces, as well as a semi-infinite geometry with only one surface. We have analyzed the role of crystal symmetries for each case. Surface states are unaltered when a single substitutional impurity is used in a semi-infinite system, yielding a band structure with Dirac cones as well as spin-polarized impurity states. In contrast, a slab geometry even with 16 atomic layers presents coupling of states at opposite surfaces, which produces a small gap in the Dirac cones. Our calculations allow us to isolate the contribution of either the impurity or the first atomic layers of the surface to the electronic structure of the system, which is specially useful since we consider different impurity positions. Impurity states show an interesting alternating behavior depending on the location of the impurity inside the material, which can be regarded as an even-odd effect. 

A continuum Hamiltonian is also used to describe the effect of doping on the Dirac cone, which shifts downwards. This result coincides qualitatively with our DFT calculations and allows us to predict the behavior of the system when the impurity is placed deeper in the slab. As for the hybridization between the topologically-protected surface states and the impurity states, a simplified version of the aforementioned Hamiltonian is used, showing a diminishing resonance as the impurity moves inwards. 

The $n$-doped character of SnTe:Sb could make difficult to separate the contribution to the electronic properties of surfaces and bulk states. However, angle-resolved photoemission spectroscopy (ARPES) enables revealing the signatures of surface states (i.e. Dirac cone) even if the Fermi level lies well above the Dirac point, as was demonstrated by Chen \emph{et al.} in the Bi$_2$Te$_3$ samples~\cite{Chen2010}. We are then confident that our predictions could be observed in ARPES experiments.

In conclusion, SnTe, the first discovered TCI, is an excellent platform to explore the interplay of symmetry and topology in this class of Dirac matter. A simple perturbation such as a substitutional impurity is shown to give rise to non-trivial spin textures. Besides its fundamental interest, tuning the Dirac cones of topological insulators can be of interest for transport and spintronic applications.

\appendix

\section*{Appendix A}
\subsection*{Additional DFT results} \label{App:DFT}
A more detailed analysis of the orbital character and spin texture of pristine and impurity-doped SnTe is illustrated by the following figures. Figure~\ref{figA0} (a) displays the band structure along with the projected contributions of all chemical species (green for Sn, blue for Te and orange for Sb) for the slab structure. The band inversion at the $\bar{\Gamma}$ point is observed for both pristine and doped structures. The impurity states hybridize with the upwards Sn-like branch of the Dirac cone. Panel (b) confirms the band inversion in bulk pristine and Sb-doped SnTe.

Figure~\ref{figA1} depicts $\mathrm{PDOS}(k_x,k_y,E)$ and spin texture maps of SnTe above the Dirac point. Two pairs of Dirac cones appear, which are folded into a $\times$-shaped fourfold pattern at the $\Gamma$ point for the c$(4\times 4)$ supercell. The $M_z({\bm k},E)$ component is omitted due to the absence of out-of-plane spin texture.

Figure~\ref{figA2} shows the persistence of the pristine SnTe spin texture when impurities are included. The helicity is unaltered independently of the impurity position, as mentioned in figure~\ref{fig3}. Additionally, it can be seen in figure~\ref{fig4} that the spin texture changes sign upon moving above or below the Dirac point.

\begin{figure*}
    \centering
    \includegraphics[width=0.6\textwidth]{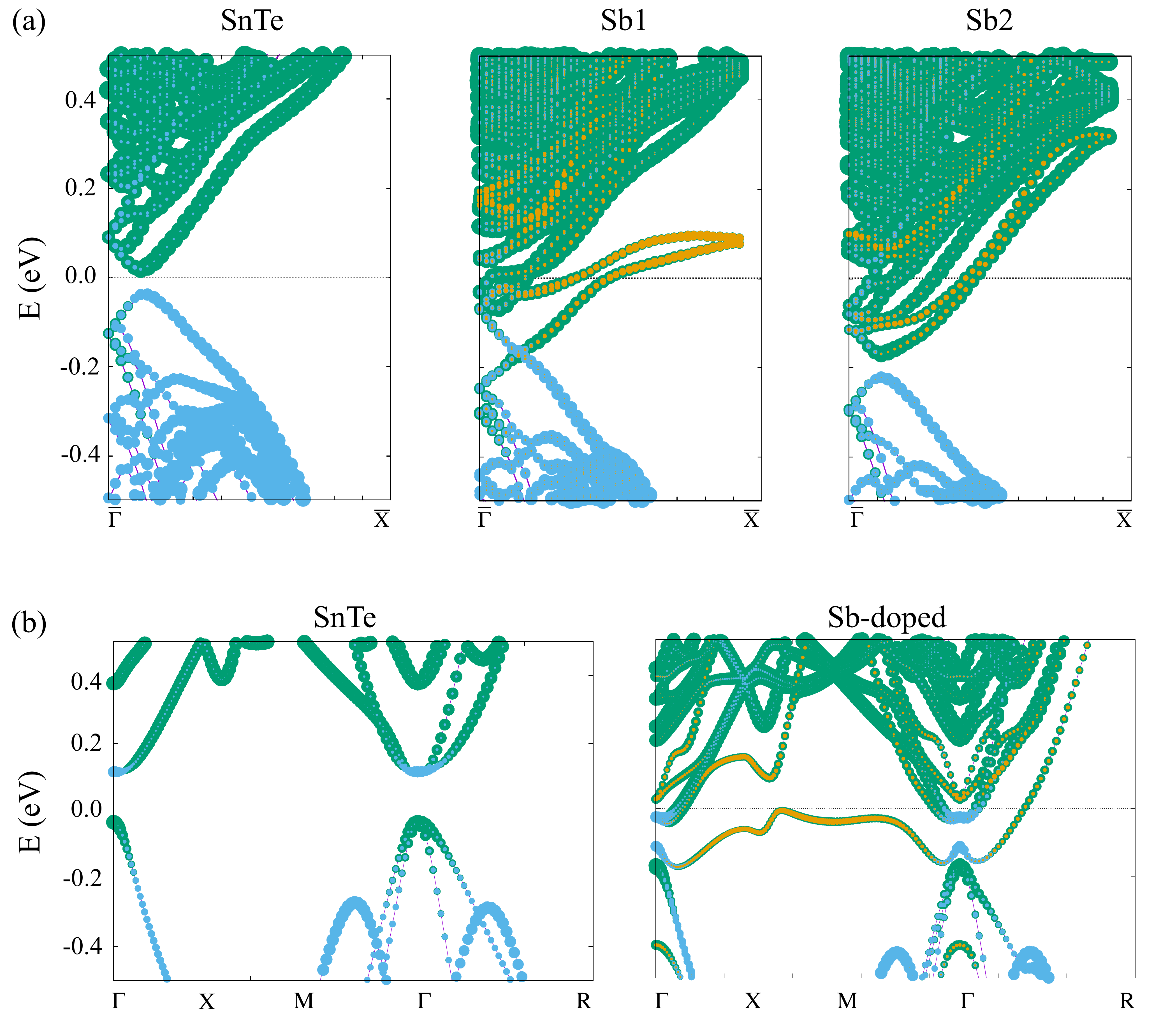}
\caption{(a) Electronic structure of the 16-layer slabs for the undoped, Sb1 and Sb2 structures, along with the projections on Sn (green), Te (blue) and Sb (orange) $p$-orbitals. (b)~Bulk band structure of the pristine and doped supercell. The color scheme is the same as in panel (a).}
\label{figA0}
\end{figure*}

\begin{figure*}[]
\centering
\includegraphics[width=0.6\textwidth]{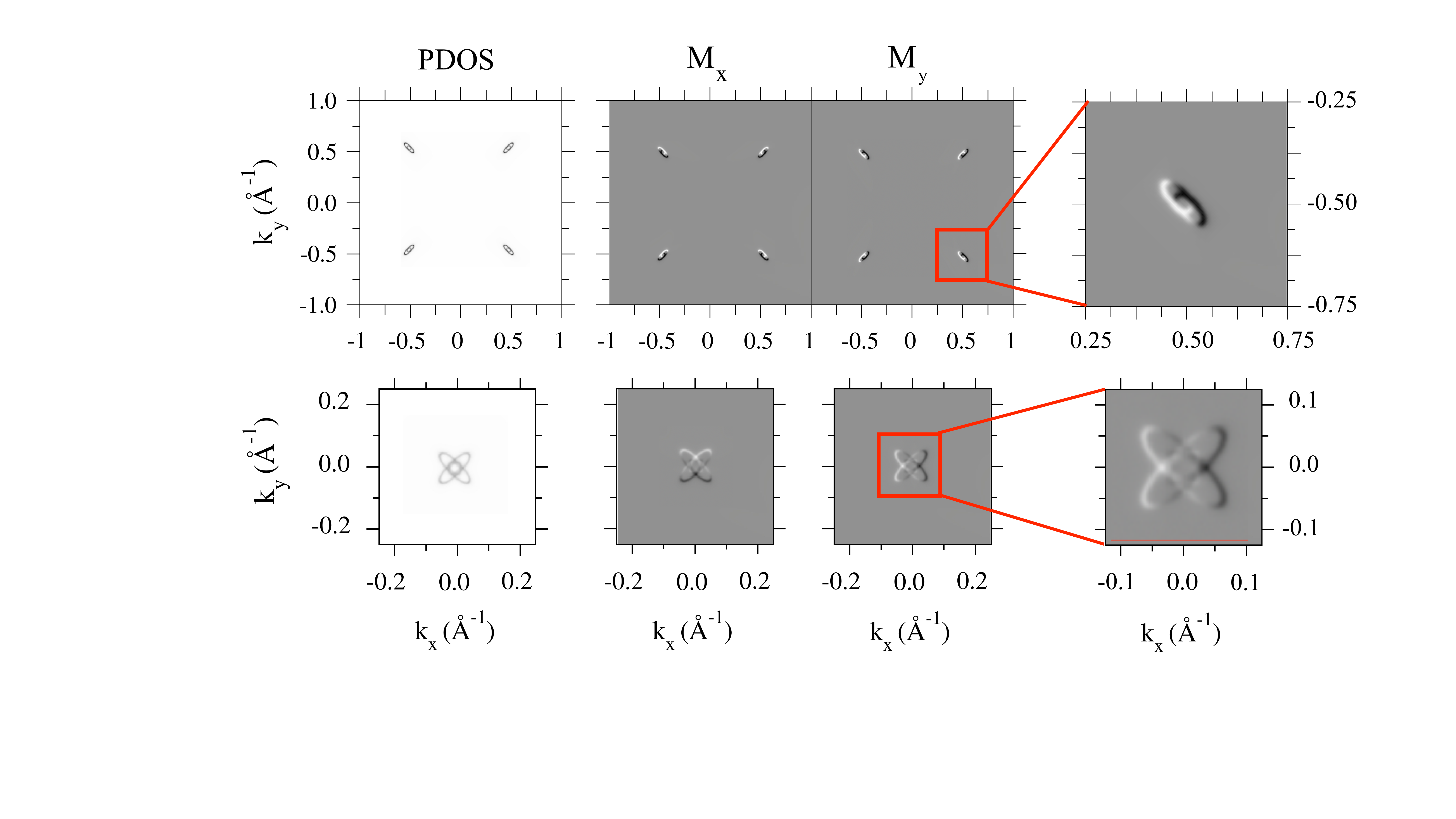}
\caption{PDOS($k_x$,$k_y$) and maps of the magnetization components $M_x(k_x,k_y)$ and $M_y(k_x,k_y)$ of pristine SnTe at 0.05 eV. Top panel shows the unfolded $1 \times 1$ unit cell, demonstrating the appearance of four helical edge states. Bottom panel shows the c$(4\times 4)$ folded unit cell; the previous helical states arrange around $\Gamma$ in a fourfold pattern. White and black shades account for positive and negative values of $M_i$, respectively.}
\label{figA1}
\end{figure*}

\begin{figure*}[ht!]
    \centering
    \includegraphics[width=0.6\textwidth]{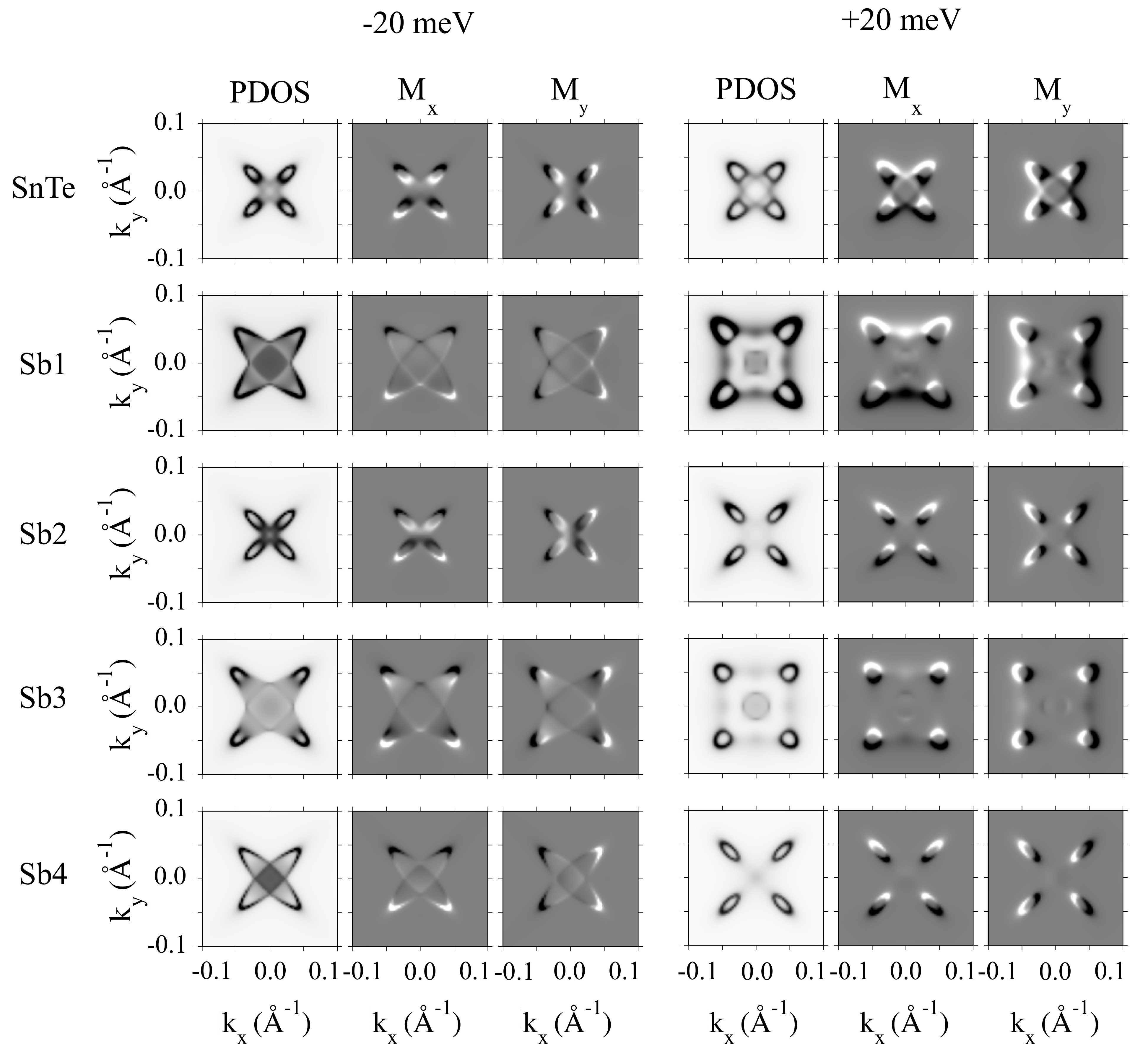}
    \caption{PDOS($k_x$,$k_y$) and maps of the magnetization components $M_x(k_x,k_y)$ and $M_y(k_x,k_y)$ at an energy $20\,$meV below (left panels) and above (right panels) the Dirac point. White and black shades account for positive and negative values of $M_i$, respectively.}
    \label{figA2}
\end{figure*}

\section*{Appendix B}
\subsection*{Surface states in the quadratic Dirac equation} \label{App:SurfStates}

The surface states used in the analytic calculations are obtained considering a semi-infinite slab extended for $z>0$ and with periodic boundary conditions in the $x$- and $y$-directions. DFT calculations use a periodical supercell of size $L_x\times L_y$; in order to match the same conditions of the \textit{ab-initio} structure we consider periodic boundary conditions in $x$- and $y$-directions while the $z$ coordinate is unbounded in performing the integrals of equation~\eqref{PT:eq:PertEn}.

Due to the breaking of translational symmetry, we solve the Hamiltonian \eqref{PT:eq:DeltaH} with $k_z \to - i \partial_z$, using the \textit{ansatz} of the form
\begin{equation} \label{SS:eq:Ansatz}
\psi \sim e^{-\la z}\exp{[i(k_x x + k_y y)]} \Phi~;
\end{equation}
in other words, we assumed plane waves in the in-plane directions, an exponential decay form in the $z$- direction and a constant spinorial part $\Phi$.

We found that two decay lengths satisfy the eigenvalue equation

\begin{equation} \label{SS:eq:la_i}
        \la_{1,2} = \frac{v}{2 \abs{B}}\left(  1 \pm \sqrt{1- \frac{4m\abs{B}}{v^2}+\frac{4 B^2 k_\parallel^2}{v^2}}\right)~.
\end{equation}

By combining solutions of the form \eqref{SS:eq:Ansatz} in order to fulfill the hard-wall boundary conditions at $z=0$, we obtain two surface states with linearly dispersive energy given by $ E_\pm = \epsilon  \pm v k_\parallel \sign(B)$ with wave functions 
\begin{align}
        \psi_+ = A_s (e^{-\la_1 z} - e^{-\la_2 z}) \exp{[i(k_x x + k_y y)]} \Phi^0_+~,\\
        \psi_- = A_s (e^{-\la_1 z} - e^{-\la_2 z}) \exp{[i(k_x x + k_y y)]} \Phi^0_-~,
\end{align}
where $\Phi^0_\pm$ are space-independent normalized spinors and the normalization constant depends on the momenta and model parameters
\begin{equation} \label{SS:eq:As}
        A_s = \frac{1}{\sqrt{L_x L_y}} \sqrt{\frac{2 \la_1 \la_2 (\la_1+\la_2)}{(\la_1-\la_2)^2}}~.
\end{equation}
The numerical values of the parameters have been obtained by fitting to DFT data and are reported in Table~\ref{PT:tab:params}.
\begin{table}[ht]
\centering{}%
\begin{tabular}{l}
\hline
\hline
\textbf{SnTe}  \\ 
\hline
\hline
         $\epsilon = \SI{-0.065}{\eV}$ \\
         $m = \SI{0.07}{\eV}$\\
         $v= \SI{1.591}{\eV \angstrom}$ \\
         $B = \SI{2.74}{\eV \angstrom^{2}}$\\ 
\hline
\hline
\end{tabular}
    \caption{Fitting parameters of the continuum model to DFT data.}
    \label{PT:tab:params}
\end{table}

In equation~\eqref{PT:eq:DeltaH}, replacing the electron free mass by the effective mass and the vacuum dielectric constant by that of the medium, the following relation for the effective Rydberg and Bohr radius can be easily obtained
\begin{equation}
    R_y^* a_B^* = {R_y a_B}/{\varepsilon_r}~,
    \label{PT:effectiveunits}
\end{equation}
where $R_y$ and $a_B$ are the well-known atomic Rydberg energy and Bohr radius and $\varepsilon_r$ is the relative dielectric permittivity. The value of the permittivity in SnTe has been tabulated in multiple references~\cite{Riedl1965Dielectric, Cardona1965Dielectric} for a wide range of temperatures. In our DFT simulations, as already mentioned in the previous sections, we are in a room-temperature regime. Hence, the tabulated value of $\varepsilon \approx 40$ at room temperature 
is the more convenient~\cite{Riedl1965Dielectric, adachi1999}.

\subsection*{Green's function in the absence of impurity} \label{App:GFA}

When describing the envelope function ${\bm\chi}({\bm r})$, the momentum perpendicular to the growth direction ${\bm k}_{\parallel}$ is conserved and the envelope function can be factorized as ${\bm\chi}({\bm r})=\widetilde{\bm\chi}(z)\exp(i\,{\bm r}_{\parallel}\cdot{\bm k}_{\parallel})$. It is understood that the subscript $\parallel$ in a vector indicates the nullification of its $z$--component, namely ${\bm r}_{\parallel}=(x,y,0)$ and ${\bm k}_{\parallel}=(k_x,k_y,0)$.  
%

The Green's function for the unperturbed problem ($\lambda=0$) satisfies the following equation
\begin{multline}
\left[ E\mathbb{1}_4+i\,\alpha_z\,\frac{\partial\phantom{z}}{\partial z}+i\,{\bm \alpha}_\parallel\cdot
\nabla_\parallel-\beta \sign(z) \right] \!
G_0({\bm r},{\bm r}^{\prime};E)\\ = \delta({\bm r}-{\bm r}^{\prime})\mathbb{1}_4\ .
\label{eq:GFA-04}
\end{multline}
The Green's function $G_0({\bm r},{\bm r}^{\prime};E)=\langle {\bm r}|\widehat{G}_0(E)|{\bm r}^{\prime}\rangle$ is the matrix element of the unperturbed resolvent $\widehat{G}_0(E)$ in coordinate representation. Let $\widehat{G}(E)$ be the resolvent of the Hamiltonian $\widehat{H}_0+\widehat{V}$, where $\widehat{V}=|\omega\rangle \lambda \langle \omega |$ is the operator associated to the non-local separable potential~(\ref{eq:GFA-03}). $\widehat{G}(E)$ and $\widehat{G}_0(E)$ are related by the well-known equation $\widehat{G}(E)=\widehat{G}_0(E)+\widehat{G}_0(E)\,\widehat{T}(E)\,\widehat{G}_0(E)$ (see Ref.~\citenum{Economou06}), where
\begin{equation}
\widehat{T}(E)=\left[\mathbb{1}-\widehat{V}\widehat{G}_0(E)\right]^{-1}
\widehat{V}\ ,
\label{eq:GFA-05}
\end{equation}
is the transition operator. A pole of $\widehat{T}(E)$ at a real (complex) energy $E$ corresponds to a bound (resonance state). After some algebra we get the following transcendental equation for the poles of the transition operator $\widehat{T}(E)$
\begin{multline}
\det \left[
\mathbb{1}_4 - \lambda \int \mathrm{d}^3{\bm r}\int \mathrm{d}^3{\bm r}^{\prime}G_0({\bm r},{\bm r}^{\prime};E)
\right. \\ \left. \phantom{\int}\times \omega({\bm r}-{\bm r}_0)\omega({\bm r}^{\prime}-{\bm r}_0) \right]=0\ ,
\label{eq:GFA-06}
\end{multline}
where $G_0({\bm r},{\bm r}^{\prime};E)$ can be expanded as
\begin{multline}
G_0({\bm r},{\bm r}^{\prime};E)=\frac{1}{4\pi^2}\int
\mathrm{d}^2{\bm k}_{\parallel}\,G_0(z,z^{\prime};{\bm k}_{\parallel};E)\\
\times \exp\left[i{\bm k}_{\parallel}\cdot\left({\bm r}_{\parallel}-{\bm
r}_{\parallel}^{\prime}\right)\right]\ .
\label{eq:GFA-07}
\end{multline}
with
\begin{align}
G_0(z,z^{\prime};&{\bm k}_{\parallel};E)=\,\frac{e^{-\kappa|z-z^{\prime}|}}{2\kappa} \nonumber \\
&\times 
\big[i\kappa\alpha_z \,\sign (z-z^{\prime}) + \beta\sign(z)+E\mathbb{1}_4\big] \nonumber\\
&+\frac{e^{-\kappa(|z|+|z^{\prime}|)}}{2\kappa(E^2-k_\parallel^2)} \left(i\alpha_z\beta+\frac{1}{\kappa}\,\mathbb{1}_4\right) \nonumber \\
&+ \big[i\kappa\alpha_z \,\sign(z) +\beta\sign(z)+E\mathbb{1}_4\big]\ ,
\label{eq:GFA-08}
\end{align}
and $\kappa^2=1+k_\parallel^2-E^2$.

The proper Green's function of the unperturbed band-inverted heterojunction can be obtained as follows. We introduce~(\ref{eq:GFA-07}) in equation~(\ref{eq:GFA-04}) to obtain
\begin{gather}
\left[ E\mathbb{1}_4-\widehat{h}_0(z) \right]
G_0(z,z^{\prime};{\bm k}_{\parallel};E)=\delta(z-z^{\prime})\mathbb{1}_4\ , 
\nonumber\\
\widehat{h}_0(z) \equiv -i\,\alpha_z\,\frac{\partial\phantom{z}}{\partial z}+{\bm \alpha}_\parallel\cdot {\bm k}_\parallel+\beta \sign(z)\ .
\label{eq:GFA-01APP}
\end{gather}
The resolvent of $\widehat{h}_0$ is $\widehat{G}_0(E)=(E\mathbb{1}_4-\widehat{h}_0)^{-1}$. We now define the auxiliary resolvent $\widehat{g}_0(E)=(E^2\mathbb{1}_4-\widehat{h}_0^2)^{-1}$ so that $\widehat{G}_0(E)=(E\mathbb{1}_4+\widehat{h}_0)\,\widehat{g}_0(E)$. Therefore, once $\widehat{g}_0(E)$ is known, we can easily obtain $\widehat{G}_0(E)$. In coordinate representation
\begin{multline}
\left[\left(\frac{\partial^2\phantom{z^2}}{\partial z^2}-\kappa^2\right)\mathbb{1}_4-2i\beta\alpha_z\delta(z)\right]
g_0(z,z^{\prime};{\bm k}_{\parallel};E)\\ = \delta(z-z^{\prime})\mathbb{1}_4\ .
\label{eq:GFA-02APP}    
\end{multline}
This equation can be easily solved regarding the term $-2i\beta\alpha_x\delta(x)$ as an interaction potential and then using the Dyson equation. In doing so we obtain
\begin{multline}
g_0(z,z^{\prime};{\bm k}_{\parallel};E)=-\frac{1}{2\kappa}\,e^{-\kappa|z-z^{\prime}|}\mathbb{1}_4+
\frac{e^{-\kappa(|z|+|z^{\prime}|)}}{2\left(E^2-k_\parallel^2\right)}
\\ \times \left(i\beta\alpha_z+\frac{1}{\kappa}\,\mathbb{1}_4\right)\ .
\label{eq:GFA-03APP}    
\end{multline}
Finally, recalling that $\widehat{G}_0(E)=(E+\widehat{h}_0)\,\widehat{g}_0(E)$, we get equation~(\ref{eq:GFA-08}).

Once the basic equations are presented, we now need to take a particular shape function to perform the calculation. In what follows we consider
\begin{equation}
\omega({\bm r}-{\bm r_0})=f(z-z_0)\delta({\bm r}_{\parallel})\ ,
\label{eq:GFA-09}
\end{equation}
where $f(z)$ is a top-hat function of width $L$ and height $1/L$, centered at $z_0$ and approaching the $\delta$-function limit ($L\ll 1$). The transcendental equation~(\ref{eq:GFA-06}) for the poles reduces to
\begin{subequations}
\begin{gather}
\det[\mathbb{1}_4-M(E)]=0\ ,
\label{eq:GFA-10a}\\
\begin{split}
M(E)\equiv \frac{\lambda}{4\pi^2}\int \! \mathrm{d}^2{\bm k}_{\parallel}\!\!\int \!\mathrm{d}z \!\!\int \! \mathrm{d}z^{\prime} 
f(z-z_0) f(z^{\prime}-z_0) \\ \times G_0(z,z^{\prime};{\bm k}_{\parallel};E)\ .
\end{split}
\label{eq:GFA-10b}
\end{gather}
\label{eq:GFA-10}
\end{subequations}
We have numerically solved equation~(\ref{eq:GFA-10}a) considering a finite bandwidth. This is equivalent to introduce an upper cutoff for the in-plane momentum, $k_\mathrm{c}$. In this way, we prevent divergences when taking the limit $L\to 0$, i.e. $f(z-z_0)\to \delta(z-z_0)$.

\section*{Conflicts of interest}
There are no conflicts to declare.\\

\section*{Acknowledgements}
We thank Gloria Platero for generously sharing her computational resources and Sergio Bravo and \'{A}lvaro D\'{\i}az-Fern\'{a}ndez for helpful discussions. This work was supported by Ministerio de Econom\'{i}a y Competitividad, Spanish MCIN and AEI and the European Union under Grants PGC2018-097018-B-I00 (MCIN/AEI/FEDER, UE), PRE2019-088874 funded by MCIN/AEI/ 10.13039/501100011033 and by “ESF Investing in your future”, and PID2019-106820RB-C21.

\balance


\bibliography{references}

\bibliographystyle{iopart-num.bst}

\end{document}